\documentclass[twocolumn,aps,prb,superscriptaddress]{revtex4-2}
\usepackage{amsmath}
\usepackage{amsfonts}
\usepackage{amssymb}
\usepackage{graphicx}
\usepackage{bm}
\usepackage{bbold}
\usepackage{color}
\usepackage{tabularx}
\usepackage{url}
\usepackage{booktabs}
\usepackage{hyperref}
\hypersetup{colorlinks,allcolors=blue}

\begin{document}

\title{Quantum phase transition driven by competing intralayer and interlayer hopping in bilayer nickelates}
\author{Xiaoyu Zhu}
\email{zxy00yxz@gmail.com}
\affiliation{International Center for Quantum Design of Functional Materials (ICQD), Hefei National Research Center for Physical Sciences at the Microscale, University of Science and Technology of China, Hefei, Anhui 230026, China}

\author{Wei Qin}
\affiliation{Department of Physics, University of Science and Technology of China, Hefei, Anhui 230026, China}

\author{Ping Cui}
\email{cuipg@ustc.edu.cn}
\affiliation{International Center for Quantum Design of Functional Materials (ICQD), Hefei National Research Center for Physical Sciences at the Microscale, University of Science and Technology of China, Hefei, Anhui 230026, China}
\affiliation{Hefei National Laboratory, University of Science and Technology of China, Hefei, Anhui 230088, China}

\author{Zhenyu Zhang}
\affiliation{International Center for Quantum Design of Functional Materials (ICQD), Hefei National Research Center for Physical Sciences at the Microscale, University of Science and Technology of China, Hefei, Anhui 230026, China}
\affiliation{Hefei National Laboratory, University of Science and Technology of China, Hefei, Anhui 230088, China}
\date{\today}

\begin{abstract}
  
  Bilayer nickelates exhibit high-temperature superconductivity under proper hydrostatic pressure or epitaxial strain, signifying the emergence of quantum phase transitions whose physical mechanisms remain unclear. Using a minimal bilayer Hubbard model incorporating only the Ni-$d_{3z^2-r^2}$ orbitals, we demonstrate that a phase transition naturally arises from tuning the ratio of intralayer to interlayer hopping amplitudes. The transition point separates regimes with a rich interplay between superconducting and density-wave orders. In the regime of weaker intralayer hopping, the ground state is characterized by quasi-long-range spin-density-wave order. As the intralayer hopping increases, the system undergoes a transition marked by the opening of a finite spin gap and the disappearance of spin-density-wave order. Meanwhile, superconductivity is dramatically enhanced, accompanied by the emergence of quasi-long-range charge-density-wave order, indicating that the system enters Luther-Emery phase. This quantum phase transition, driven by the competition between intralayer and interlayer hopping, provides a plausible microscopic explanation for the experimentally observed correlation between the superconducting transition temperature and ratio of out-of-plane to in-plane lattice constants. Our findings reveal a possible link between the suppression of spin-density-wave order and the prominence of superconducting order, which may assist future efforts to optimize experimental conditions for further enhancing superconductivity in bilayer nickelates.
  
\end{abstract}

\maketitle

\section{Introduction}

Quantum phase transitions take place when the ground state of a system at zero temperature experiences qualitative changes driven by non-thermal parameters \cite{sachdevQuantum2011,vojtaQuantum2003}. In strongly correlated superconductors such as cuprates, the doping-induced quantum phase transition provides important clues for understanding the pairing mechanism and anomalous states, including strange metal and pseudogap phase \cite{sachdevColloquium2003,andoQuantum2004,balakirevQuantum2009}. The recent discovery of pressure-induced superconductivity in La$_3$Ni$_2$O$_7$ has established the Ruddlesden-Popper bilayer nickelates as another high-temperature superconducting platform \cite{sunSignatures2023,houEmergence2023,wangBulk2024} that likely displays strong electronic correlations \cite{zhangHightemperature2024,liuElectronic2024,yangOrbitaldependent2024}. Here, the bulk samples exhibit signatures of density-wave orders at ambient pressure \cite{liuEvidence2022,chenElectronic2024,zhaoPressureenhanced2025,chenEvidence2024,mengDensitywavelike2024,khasanovPressure2025,plokhikhUnraveling2025}, but superconductivity emerges when the applied pressure reaches the order of 10 GPa. Thin-film bilayer nickelates can superconduct without extra hydrostatic pressure, yet still require external tuning via substrate-induced epitaxial strain \cite{koSignatures2025,zhouAmbientpressure2025}. These observations suggest a putative quantum phase transition, the close inspection of which may clarify how superconductivity emerges from competing orders in bilayer nickelates \cite{zhangEmergent2024,jiangIntertwined2025,leonovElectronic2024}.

It is widely believed that the Ni-$d_{3z^2-r^2}$ and Ni-$d_{x^2-y^2}$ orbitals near the Fermi level are relevant to superconductivity in bilayer nickelates \cite{sunSignatures2023,luoBilayer2023,zhangStructural2024,shenEffective2023,yangPossible2023}. Given their markedly distinct effective intralayer and interlayer hopping amplitudes, the two orbitals may play qualitatively different roles \cite{jiangPressure2024,sakakibaraPossible2024,kanekoPair2024,luoHigh$T_c$2024,yangStrong2024,zhangStrong2024,wangHighly2025,zheng$s_pm$wave2025,maierInterlayer2025,fanSuperconductivity2024}. Specifically, the $d_{3z^2-r^2}$ electrons can form interlayer singlet pairs in this strongly correlated system, facilitated by the strong interlayer hopping, but the weak intralayer hopping limits their phase coherence. In contrast, while the $d_{x^2-y^2}$ orbital exhibits strong intralayer hopping, the pairing is weak due to the negligible interlayer hopping and low electron filling. This naturally raises the question of how the two orbitals contribute to the high-temperature superconductivity, which demands both strong pairing and robust phase coherence. Some theories suggest that hybridization of the two orbitals enhances the phase coherence of Cooper pairs formed by $d_{3z^2-r^2}$ electrons \cite{yangInterlayer2023,qinHigh$T_c$2023}, and other studies propose that Hund's coupling plays a critical role in the pairing of $d_{x^2-y^2}$ electrons \cite{ohTypeII2023,luInterlayerCouplingDriven2024,quBilayer2024,luInterplay2024,jiStrongCouplingLimit2025}. To clarify their individual roles in the emergence of superconductivity under external tuning, one possible approach is to examine each orbital separately while treating the interorbital coupling as a tunable parameter. Exploring superconductivity and the associated quantum phase transition in this effective single-orbital model may offer key insights into the pairing mechanism in bilayer nickelates.

\begin{figure}[ht]
  \includegraphics[width=0.48\textwidth]{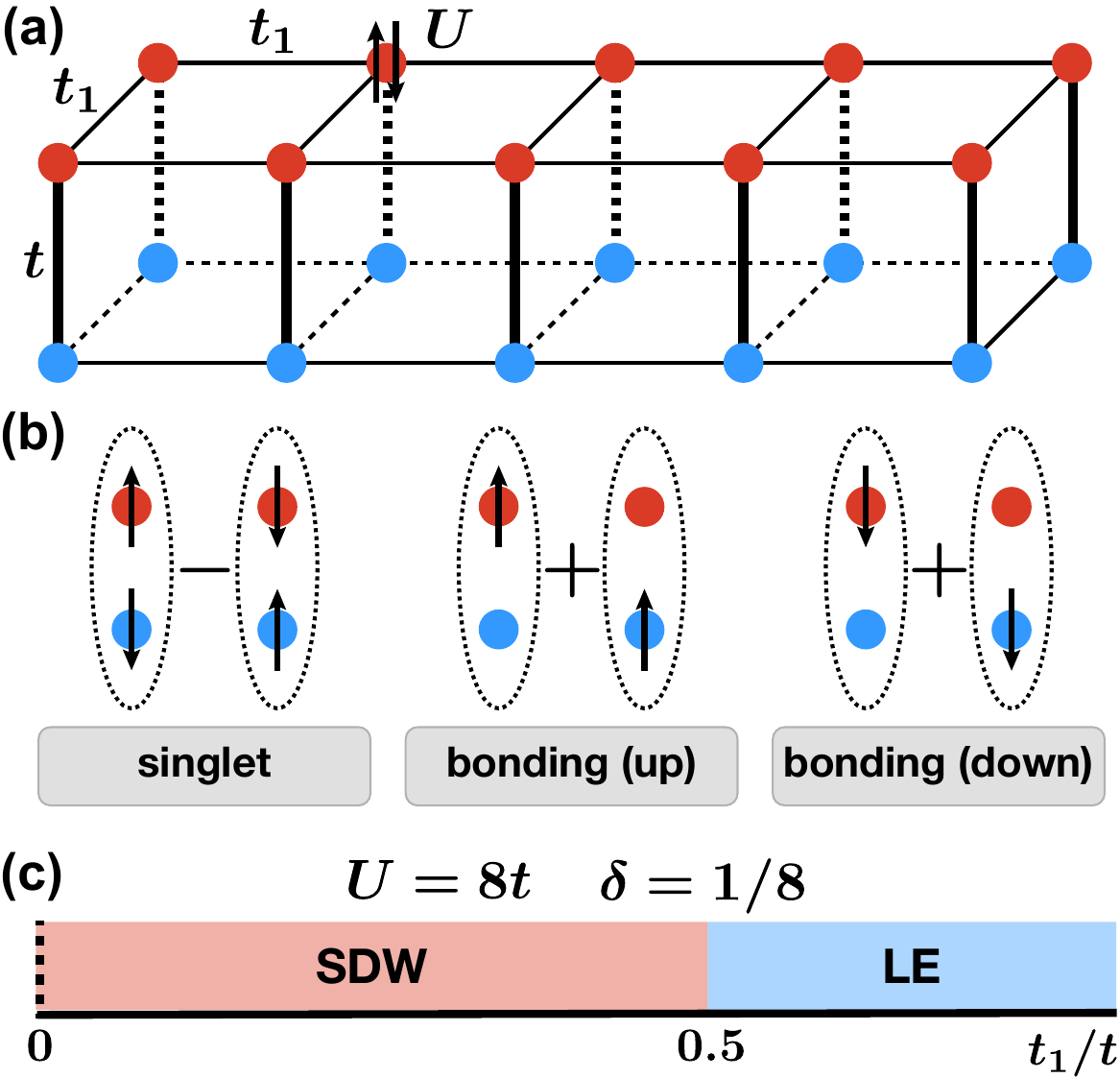}
  \caption{(a) Schematic of the bilayer Hubbard model. Circles of different colors indicate electron sites in different layers. The strong interlayer hopping $t$ is indicated by thick lines while the weak intralayer hopping $t_1$ is shown in thin lines. (b) Three types of dimers relevant to ground states in the weak $t_1$ regime. (c) Two distinct phases separated by a quantum phase transition at $t_1\approx 0.5t$, calculated for onsite Hubbard interaction $U = 8t$ and doping $\delta = 1/8$. The dashed line indicates the dimer limit ($t_1=0$), where the multiply degenerate ground states only involve the three types of dimers in (b). SDW is the shorthand for spin density wave, and LE denotes Luther-Emery liquid, which features competing superconductivity and charge density wave.}\label{fig:model}
\end{figure}

In this article, we focus on the $d_{3z^2-r^2}$ orbital, as previous studies have suggested a link between its bonding band and the emergence of superconductivity \cite{sunSignatures2023,luoBilayer2023,yangPossible2023,sakakibaraPossible2024,yueCorrelated2025,liAngleresolved2025}, highlighting its possible role in the quantum phase transition. The tuning parameter we consider is the ratio of intralayer to interlayer hopping, which can be varied by pressure or strain in bilayer nickelates. We expect an increasing ratio to promote superconducting order, as enhanced intralayer hopping between $d_{3z^2-r^2}$ orbitals facilitates phase coherence among interlayer singlets. The bilayer Hubbard model \cite{scalapinoCommon2012,dagottoSuperconductivity1992,bulutNodeless1992,hetzelPairing1994,dossantosMagnetism1995,liechtenstein$s$Wave1995,kancharlaBand2007,bouadimMagnetic2008,zhaiAntiferromagnetically2009,lanataSuperconductivity2009,maierPair2011,maierEffective2019,liuHybridizationinduced2025,prasadFinitetemperature2022} then provides a minimal setting that encompasses these necessary ingredients. Our density matrix renormalization group (DMRG) \cite{whiteDensity1992,whiteDensitymatrix1993} simulations reveal a quantum phase transition that occurs when the intralayer hopping amplitude reaches approximately half that of interlayer hopping. Figure \ref{fig:model}(c) illustrates the two distinct phases, with one featuring quasi-long-range spin density wave (SDW) order, and the other, denoted by Luther-Emery (LE) phase, featuring the competition between quais-long-range superconducting (SC) order and quais-long-range charge-density-wave (CDW) order. The superconducting correlations are enhanced as the ratio of intralayer to interlayer hopping increases, consistent with recent thin-film experiments showing that the superconducting transition temperature is positively correlated with the ratio of out-of-plane to in-plane lattice constants \cite{koSignatures2025,osadaStraintuning2025}. Our results suggest that studying quantum phase transitions in bilayer nickelates may help identify competing orders and pinpoint the key factors responsible for high-temperature superconductivity in these materials.

\section{Model Hamiltonian}

As illustrated in Fig. \ref{fig:model}(a), the bilayer model we consider consists of two coupled layers with onsite Hubbard interaction $U$, interlayer hopping $t$, and nearest-neighbor intralayer hopping $t_1$. The Hamiltonian is given by
\begin{align}
  H & = -t_1\sum_{\langle \vec{r},\vec{r}' \rangle}^{\lambda,\sigma}c_{\vec{r},\lambda\sigma}^\dagger c^{\vphantom{\dagger}}_{\vec{r}',\lambda\sigma} - t \sum_{\vec{r}}^{\lambda,\sigma} (c^\dagger_{\vec{r},1\sigma} c^{\vphantom{\dagger}}_{\vec{r},2\sigma} + \text{h.c.}) \nonumber\\
   & + U \sum_{\vec{r}}^{\lambda} n_{\vec{r},\lambda\uparrow} n_{\vec{r},\lambda\downarrow},\label{eq:model}
\end{align}
where $\vec r = (x,y)$, $\lambda = 1,2$, and $\sigma = \uparrow,\downarrow$, denoting the in-plane lattice site, layer, and spin, respectively. The electron number operator is defined as $n_{\vec{r},\lambda\sigma} = c_{\vec{r},\lambda\sigma}^\dagger c^{\vphantom{\dagger}}_{\vec{r},\lambda\sigma}$. In the following, we take $U=8t$ and focus on $t_1<0.7t$, corresponding to the intermediate-to-strong coupling regime as $U$ is slightly larger than the bandwidth \cite{arovasHubbard2022,qinHubbard2022}.

\section{Dimer limit}

In the absence of intralayer hopping, the bilayer model reduces to an array of isolated dimers. Table \ref{table:dimer} lists the seven lowest energy levels of a single dimer. The fourfold-degenerate states in the first row correspond to the empty state and the three triplet states. The state in the second row is approximately an interlayer spin singlet, as $\eta$ ($=\sqrt{5}-2$) is much smaller than 1 for $U=8t$. The two degenerate states in the last row are exactly the bonding states in the non-interacting limit.

\begin{table}[ht]
  \caption{The seven lowest energy levels for a single dimer. $|\uparrow_\lambda(\downarrow_\lambda)\rangle$ denotes a spin-up (down) electron on layer $\lambda$, and $|0\rangle$ denotes the empty state. $\eta = \alpha/(1+\sqrt{1+\alpha^2})$ with $\alpha=4t/U$, and $\mathcal{A}$ is the normalization constant.}
  \label{table:dimer}
  \begin{ruledtabular}
  \begin{tabular}{cc}
    $E$ & $\psi$ \\
    \hline
    $0$ & $|0\rangle$, $|\uparrow_1 \uparrow_2\rangle$, $|\downarrow_1 \downarrow_2\rangle$, $(|\uparrow_1 \downarrow_2\rangle+|\downarrow_1 \uparrow_2\rangle)/\sqrt{2}$ \\ 
    $-2\eta t$ & $\mathcal{A}[\eta(|\uparrow_1\downarrow_1 \rangle + |\uparrow_2\downarrow_2\rangle) + (|\uparrow_1 \downarrow_2\rangle-|\downarrow_1 \uparrow_2\rangle)]$ \\
    $-t$ & $(|\uparrow_1\rangle+|\uparrow_2\rangle)/\sqrt{2}$, $(|\downarrow_1\rangle+|\downarrow_2\rangle)/\sqrt{2}$ \\
  \end{tabular}
  \end{ruledtabular}
\end{table}

At half filling, the ground state of the dimer array is unique, with each dimer residing in the singlet state. Upon hole doping, some dimers prefer the bonding states. Define the doping parameter as $\delta = N_h/N$, where $N$ and $N_h$ denote the number of lattice sites and introduced holes, respectively. For $\delta<1/2$, there are $N\delta $ dimers occupying the bonding states in the ground state, while the rest stay in the singlet state. Various arrangements of the three types of dimers displayed in Fig. \ref{fig:model}(b) give rise to highly degenerate ground states that may differ in total spin. Consequently, the system in the dimer limit supports gapless spin excitations.

At finite $\delta$, the degenerate ground states are separated from the first excited states by a gap of $2\eta t$, corresponding to an excitation that flips a singlet dimer into a triplet one. The intralayer hopping $t_1$ mixes these ground states and lifts their degeneracy. As long as $t_1$ remains sufficiently small compared to the gap $2\eta t$, the ground state can be approximated as a superposition of the three types of dimers. When $\delta$ is small, the interlayer pairing is robust due to the dominance of singlet dimers in the ground state. The $t_1$ term is expected to increase phase coherence and thereby enhance superconductivity, as previously suggested by Kuroki \emph{et al.} \cite{kurokiHightemperature2002} using the fluctuation exchange method. In the following, we investigate the emergence of superconductivity in detail, focusing on its interplay with competing orders.

\section{Finite intralayer hopping}

We consider a bilayer $N_x\times N_y$ lattice with $N_y=2$. Open boundary conditions are imposed in both directions. Using the ITensor library \cite{fishmanITensor2022}, we run DMRG simulations with various truncation errors $\epsilon$, reaching as small as $\epsilon = 1.0\times 10^{-7}$, and extrapolate the results to $\epsilon=0$ using quadratic fitting. The doping parameter is fixed at $\delta=1/8$. According to the analysis above, the ratio of bonding to singlet dimers is precisely $1/3$ in the dimer limit. Owing to the distinct charge, spin, and pairing properties of the three types of dimers, their distribution and superposition in the ground state can influence corresponding local densities and correlation functions.

\begin{figure}[t]
  \includegraphics[width = 0.48\textwidth]{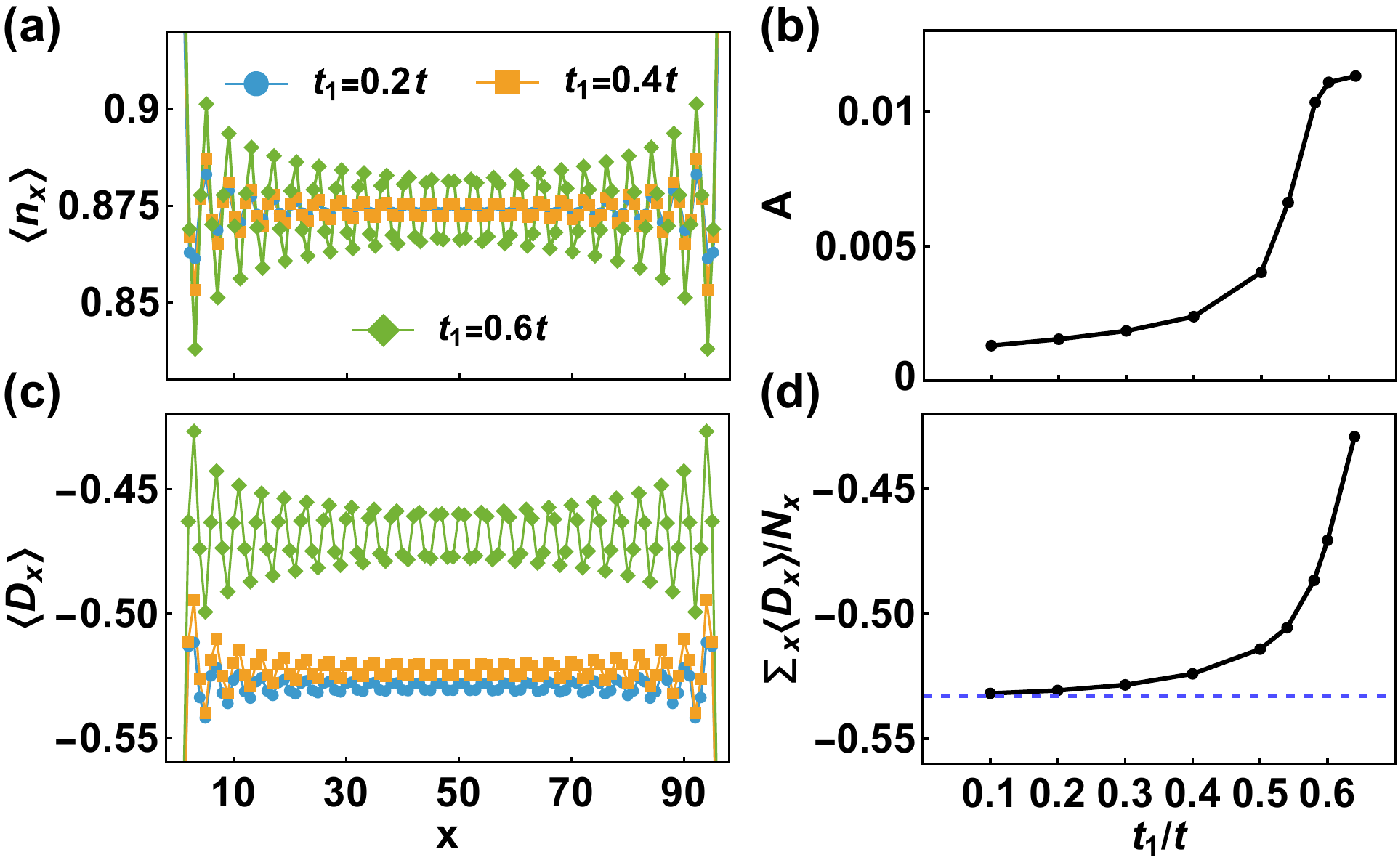}
  \caption{(a) Charge-density profiles along the $x$ direction for $t_1=0.2t$, $0.4t$, and $0.6t$. (b) Oscillation amplitude $A$ as a function of $t_1/t$. (c) Spatial profiles of spin-spin correlation $\langle D_x\rangle$ within each dimer at $t_1=0.2t$, $0.4t$, and $0.6t$. (d) Evolution of the averaged $\langle D_x \rangle$ with $t_1/t$. The blue dashed line marks the average value in the dimer limit. These plots are obtained on a bilayer $96\times 2$ lattice. }\label{fig:density}
\end{figure}

\subsection{Density oscillation}
We begin by examining the distribution of charge density along the $x$ direction, defined as $\langle n_{x,\lambda}\rangle = \frac{1}{N_y}\sum_y \langle n_{\vec r,\lambda\uparrow}+n_{\vec r,\lambda\downarrow}\rangle$. As shown in Fig. \ref{fig:density}(a), the charge density exhibits oscillations with a period of $1/(2\delta) = 4$. We have checked that the two layers display identical density distributions. To investigate how the oscillation amplitude evolves with $t_1/t$, we fit the density profile in the range $x\in [N_x/4+1,3N_x/4]$ using the ansatz $\langle n_x \rangle = n_0 + A \cos (\pi x/2-\phi)$ \cite{jiangSuperconductivity2019}. Considering the reflection symmetry, we assume the phase $\phi = \pi(N_x+1)/4$, leaving $n_0$ and $A$ as fitting parameters. Figure \ref{fig:density}(b) plots $A$ versus $t_1/t$, which exhibits a sharp increase in the oscillation amplitude around $t_1=0.5t$. As we will show later, this signals the onset of quasi-long-range CDW order.

The spin density vanishes at each site, since the continuous SU(2) symmetry cannot be spontaneously broken in this quasi-one-dimensional (1D) lattice \cite{hohenbergExistence1967,merminAbsence1966}. Nevertheless, we can investigate the distribution of spin-spin correlation within each dimer along the $x$ direction, which takes the form
\begin{equation}
  \langle D_{x}\rangle = \frac{1}{N_y}\sum_{y} \langle \vec S_{\vec r,1}\cdot \vec S_{\vec r,2}\rangle,\label{eq:rung-singlet}
\end{equation}
with $\vec S_{\vec r,\lambda}$ denoting the spin operator in layer $\lambda$. According to Table \ref{table:dimer}, $\langle D_x \rangle \approx -0.71 $ for the singlet dimer, while for the bonding dimers $\langle D_x \rangle = 0$. $\langle D_x \rangle$ also exhibits period-4 oscillations, as shown in Fig. \ref{fig:density}(c). The peaks in the charge-density profile, where the singlet dimer has larger weight, correspond to the valleys in the $\langle D_x \rangle$ profile. At $1/8$ doping, the average of $\langle D_x \rangle$ over $x$ is $-0.71*3/4+0*1/4 \approx -0.53$ in the dimer limit, and remains nearly unchanged for small $t_1/t$, as Fig. \ref{fig:density}(d) shows. This supports our earlier argument that the ground state in the weak intralayer hopping regime can be approximately described by superposition of the three types of dimers. As $t_1$ approaches $0.5t$, the average value of $\langle D_x \rangle$ rapidly deviates from its value in the dimer limit, indicating increased contributions from higher-energy dimers, such as the triplet states, in the ground-state configuration.

\subsection{Correlation functions}

\begin{figure}[t]
  \includegraphics[width = 0.48\textwidth]{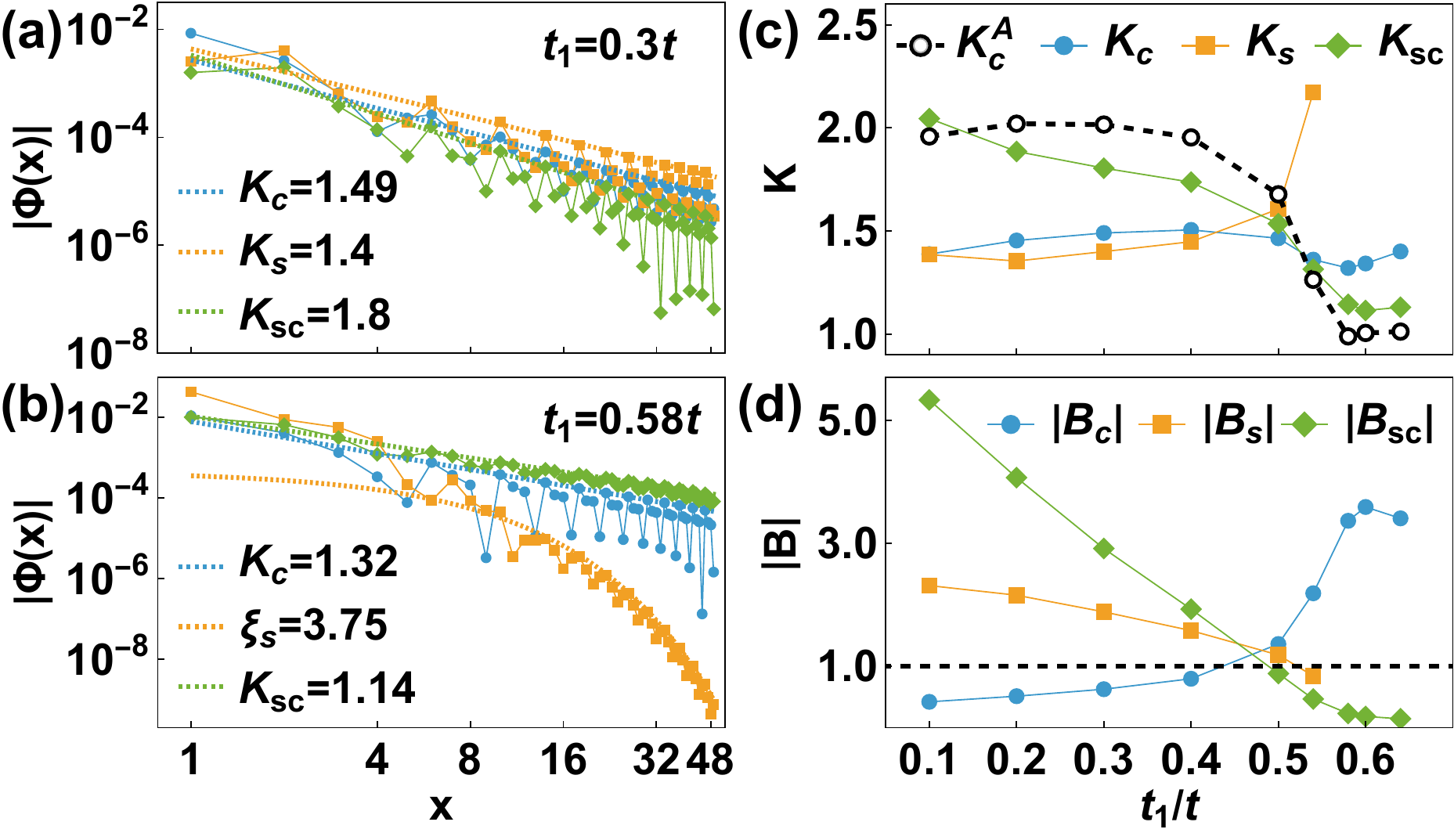}
  \caption{(a) and (b) Charge (c), spin (s), and pairing (sc) correlation functions plotted in the log-log scales. Dashed lines represent power-law or exponential fits to the envelope function $f(x)$. $K$ denotes the decay exponent of the power-law fitting, and $\xi$ is the decay length in the exponential case. (c) Dependence of decay exponents on $t_1/t$. The dashed line indicates $K_c^A$, extracted from the charge-density profile. (d) Oscillation coefficient $B$ in $P(x)$ as a function of $t_1/t$. $|B|>1$ indicates that the correlation functions change signs periodically. These results are obtained from simulations on a $96\times 2$ bilayer lattice.}\label{fig:correlation}
\end{figure}

To determine whether a phase transition occurs around $t_1=0.5t$, we further investigate the charge-charge, spin-spin, and pair-pair correlation functions \cite{jiangHigh2021}, defined respectively as
\begin{align}
  &\Phi_c(x) = \frac{1}{N_y}\sum_{y}\langle n_{\vec r_0} n_{\vec r_0 + x\vec e_x}\rangle - \langle n_{\vec r_0} \rangle \langle n_{\vec r_0 + x\vec e_x}\rangle, \nonumber \\
  &\Phi_s(x) = \frac{1}{N_y}\sum_{y}\langle \vec S_{\vec r_0} \cdot \vec S_{\vec r_0 + x\vec e_x}\rangle, \nonumber \\
  &\Phi_{sc}(x) = \frac{1}{N_y}\sum_{y}\langle \Delta_{\vec r_0}^\dagger \Delta_{\vec r_0 + x\vec e_x}^{\vphantom{\dagger}}\rangle,\label{eq:correlation}
\end{align}
where the reference site $\vec r_0=(N_x/4+1)\vec e_x+y\vec e_y$, with $\vec e_{x(y)}$ denoting the unit vector along the $x(y)$ direction, and $\Delta_{\vec r} = (c_{\vec r,1\uparrow}c_{\vec r,2\downarrow}-c_{\vec r,1\downarrow}c_{\vec r,2\uparrow})/\sqrt{2}$, representing interlayer singlet pairing. The layer index is omitted in the definitions of charge and spin correlations. Figure \ref{fig:correlation} shows the intralayer correlation functions, and we have verified that the interlayer correlations exhibit similar behavior in the real space.

In the quasi-1D system, SDW and superconductivity can at most exhibit quasi-long-range orders, characterized by algebraically decaying spin-spin and pair-pair correlation functions. Figure \ref{fig:correlation}(a) and (b) show the correlation functions for two representative values of $t_1/t$, all of them displaying oscillatory decay. We fit them using the ansatz $\Phi(x)=f(x)P(x)$, where $f(x)$ is the envelope function and $P(x)$ is a periodic function with a site period of 4. 

For $t_1=0.3t$, all the correlation functions exhibit power-law decay, and the envelope function can be fitted by $f(x) = C x^{-K}$. The corresponding susceptibility follows $\chi \sim T^{K-2}$ and diverges at zero temperature if the decay exponent satisfies $K<2$ \cite{giamarchiQuantum2003}, which suggests the emergence of truly long-ranged order in the two-dimensional (2D) limit. Here, all three decay exponents are found to be less than 2. The decay exponents of spin and charge correlations are comparable, both of which are smaller than the decay exponent of the pairing correlation, indicating that spin- or charge-related orders dominate in this regime. In the Supplemental Material \cite{supplemental}, we consider a wider system with $N_y=4$ at $t_1=0.2t$, where the conclusion holds as well.

In the case of $t_1=0.58t$, as shown in Fig. \ref{fig:correlation}(b), the pairing correlation exhibits the slowest decay, suggesting that superconducting order becomes dominant. However, the spin correlation decays exponentially, and hence its envelope function can be fitted by $f(x) = C e^{-x/\xi}$, with $\xi$ being the decay length.

The qualitatively distinct behavior of spin correlations in the two cases provides strong evidence for a phase transition in between. Figure \ref{fig:correlation}(c) shows the evolution of $K$ as a function of $t_1/t$, confirming the presence of a transition near $t_1=0.5t$. As $t_1/t$ increases past the transition point, the spin correlation becomes short-ranged, while the pairing correlation is strongly enhanced. In contrast, the decay exponent of the charge correlation, $K_c$, appears to be largely unaffected.

To identify density-wave orders, we proceed to examine the periodic part $P(x)$, which in general consists of both uniform and oscillatory components. We fit them using the ansatz $P(x) = 1 + B \cos (\pi x/2 + \theta)$. When $|B|>1$, $P(x)$ and the correlation function $\Phi(x)$ change signs periodically, signaling the presence of density-wave orders. As demonstrated in Fig. \ref{fig:correlation}(d), quasi-long-range SDW order develops and dominates for $t_1<0.5t$. In this regime, pairing correlations also exhibit sign oscillations, indicating that the oscillatory part dominates over the uniform component. After the transition, pairing correlations no longer change signs, revealing the dominance of uniform superconductivity. Although the charge correlations decay algebraically, signs of $P_c(x)$ remain constant until the transition point is approached, from which we conclude that quasi-long-range CDW order emerges only for $t_1>0.5t$.

We can also investigate CDW by fitting the scaling of charge-density oscillation amplitude with system size to the relation $A\sim N_x^{-K_c^A/2}$ \cite{whiteFriedel2002,jiangSuperconductivity2019}. In the presence of quasi-long-range CDW order, the decay exponent $K_c^A$ should be close to $K_c$ and noticeably smaller than 2. As shown in Fig. \ref{fig:correlation}(c), $K_c^A$ is considerably larger than $K_c$ and nearly equals 2 in the weak intralayer hopping regime, confirming the absence of CDW. This is further corroborated by the decay exponents extracted using Friedel oscillation formula and by the much weaker oscillation observed in a $N_y=4$ lattice, as presented in Supplemental Material \cite{supplemental}. The sharpe decrease in $K_c^A$ across the transition indicates the emergence of CDW order, consistent with the rapid enhancement of charge-density oscillations shown in Fig. \ref{fig:density}(b). However, unlike $K_c$, $K_c^A$ is slightly smaller than $K_{sc}$, suggesting a strong competition between CDW and superconductivity, a characteristic feature of Luther-Emery liquid \cite{lutherBackward1974}, as also observed in other Hubbard and $t$-$J$ models \cite{dolfiPair2015,luGroundstate2023,langePairing2024,gongRobust2021,jiangSuperconductivity2019,jiangHigh2021}.
  
\subsection{Static structure factors}

\begin{figure}
  \includegraphics[width = 0.48\textwidth]{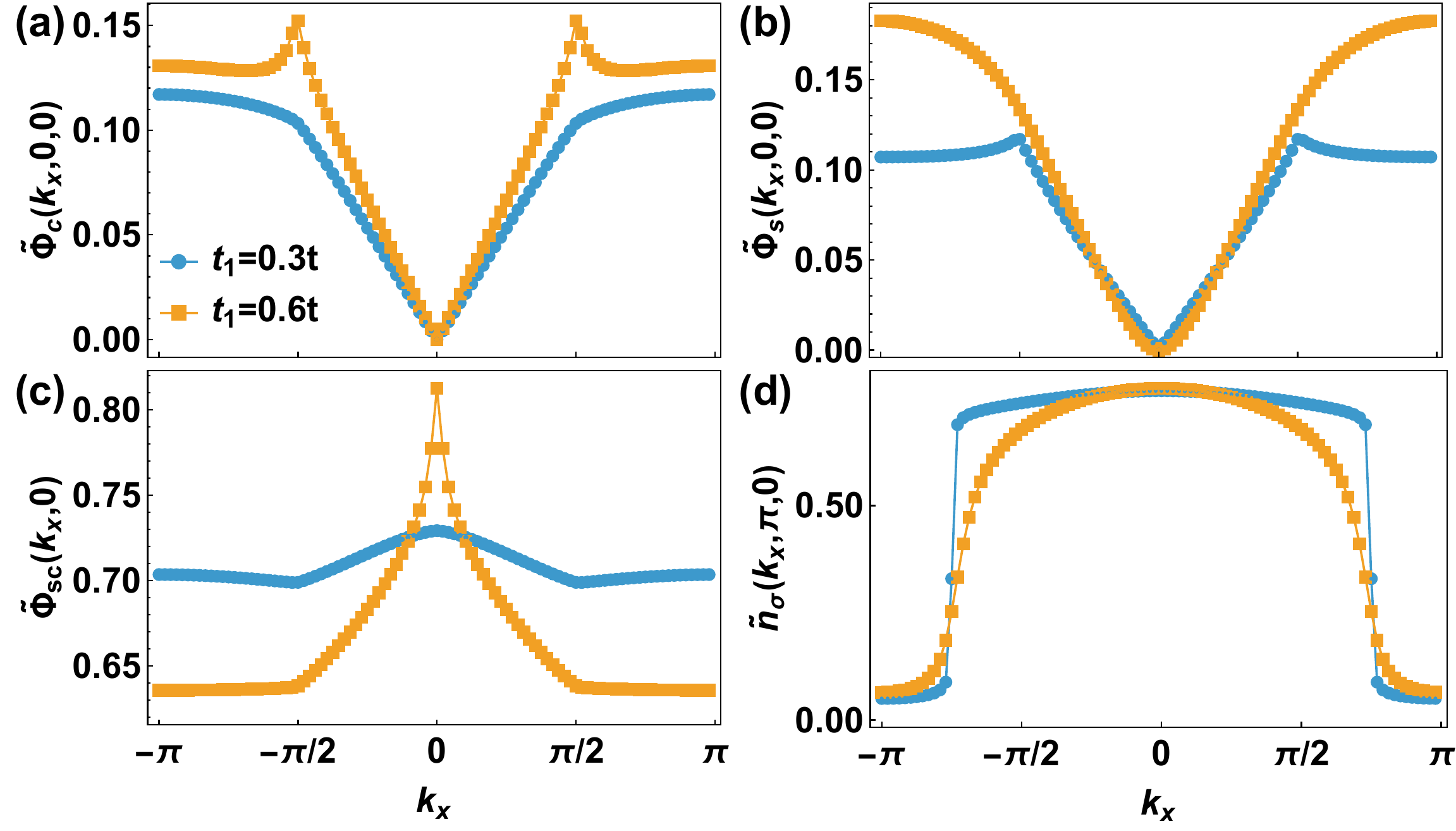}
  \caption {(a) Charge structure factor. A peak appears in the LE phase ($t_1=0.6t$) at $(\pi/2,0,0)$. (b) Spin structure factor. The peak shows only in the SDW phase ($t_1=0.3t$) at $(\pi/2,0,0)$. (c) Pair structure factor. There is a singular point in the SDW phase at $(k_x,k_y) = (\pi/2,0)$. In the LE phase, the singularity only occurs at $(0,0)$. (d) Momentum distribution function. The effective Fermi point $k_F$, where $\tilde n_{\sigma}$ becomes singular, lies at $k_x = 3\pi/4$. }\label{fig:structure}
\end{figure}

The oscillatory power-law decay of correlation functions observed in the real space could also be identified in the momentum space from the non-analyticity in static structure factors. Following the notation in Ref. \cite{maierPair2011}, we may define the static charge, spin, and pair structure factors as 
\begin{align}
  &\tilde\Phi_c(\vec k) = \frac{1}{N}\sum_{\vec r,\vec r'}^{\lambda,\lambda'}  e^{i[\vec k\cdot (\vec r - \vec r') + k_z(\lambda - \lambda')]} \langle n_{\vec r, \lambda}^c n_{\vec r',\lambda'}^c\rangle,\nonumber\\
  &\tilde\Phi_s(\vec k) = \frac{1}{N}\sum_{\vec r,\vec r'}^{\lambda,\lambda'} e^{i[\vec k\cdot (\vec r - \vec r') + k_z(\lambda - \lambda')]}\langle \vec S_{\vec r, \lambda}\cdot \vec S_{\vec r',\lambda'}\rangle,\nonumber\\
  &\tilde\Phi_{sc}(k_x,k_y) = \frac{2}{N}\sum_{\vec r,\vec r'} e^{i\vec k\cdot (\vec r - \vec r')}\langle \Delta^\dagger_{\vec r} \Delta^{\vphantom{\dagger}}_{\vec r'} \rangle
\end{align}
where $\vec k = (k_x,k_y,k_z)$ with $k_z = 0$ and $\pi$ representing the bonding and anti-bonding bands of bilayer lattice respectively, $n_{\vec r, \lambda}^c =\sum_\sigma (n_{\vec r, \lambda\sigma} - \langle n_{\vec r, \lambda\sigma} \rangle$), and $N=2N_xN_y$, being the number of lattice sites.

Our previous calculations demonstrate that quasi-long-range CDW order develops when $t_1/t>0.5$. As Fig. \ref{fig:structure} (a) shows, the charge structure factor $\tilde\Phi_c(\vec k)$ indeed exhibits a pronounced peak at $\vec k = (\pi/2,0,0)$ for $t_1=0.6t$. For spin structure factor $\tilde\Phi_s(\vec k)$ shown in Fig. \ref{fig:structure} (b), the peak appears at the same wave vector but for $t_1=0.3t$, verifying the existence of quasi-long-range SDW in the weak intralayer hopping regime. The plot of pair structure factor $\tilde\Phi_{sc}(k_x,k_y)$ in Fig. \ref{fig:structure} (c) displays a sharp peak at $(k_x,k_y) = (0,0)$ for $t_1=0.6t$, revealing the uniform superconducting order in the LE phase. It is interesting to note that a cusp also develops at $\tilde\Phi_{sc}(\pi/2,0)$ in the case $t_1=0.3t$, corroborating our earlier observation of the strong pair density modulation in the SDW phase. This hints at the existence of pair-density-wave (PDW) component, but it may not survive in the 2D limit as SDW is the dominating order.

The oscillation wave vectors of correlation functions are closely related to the effective Fermi momentum $k_F$ in the theory of Luttinger liquid, which is identified as the nonanalytical point in momentum distribution function $\tilde n_\sigma(\vec k)$, defined as
\begin{equation}
  \tilde n_\sigma(\vec k) = \frac{1}{N}\sum_{\vec r,\vec r'}^{\lambda,\lambda'} \langle c_{\vec r, \lambda\sigma}^\dagger c_{\vec r',\lambda'\sigma}^{\vphantom{\dagger}} \rangle e^{i[\vec k\cdot (\vec r - \vec r') + k_z(\lambda - \lambda')]}.
\end{equation}
As shown in Fig. \ref{fig:structure}(d), a sharp drop in $\tilde n_\sigma(\vec k)$ appears at $\vec k=(\pm 3\pi/4,\pi,0)$. Unlike in Landau's Fermi liquid, where discontinuity of $\tilde n_\sigma(\vec k)$ is expected at Fermi points, here the nonanalyticity is characterized by a diverging slope of $\tilde n_\sigma(\vec k)$ in the thermodynamic limit. Clearly, in the structure factor plot, the nonanalyticity at nonzero $k_x$ occurs exactly when $k_x=2k_F$ $(\text{mod } 2\pi)$.

\begin{figure}[t]
  \includegraphics[width = 0.48\textwidth]{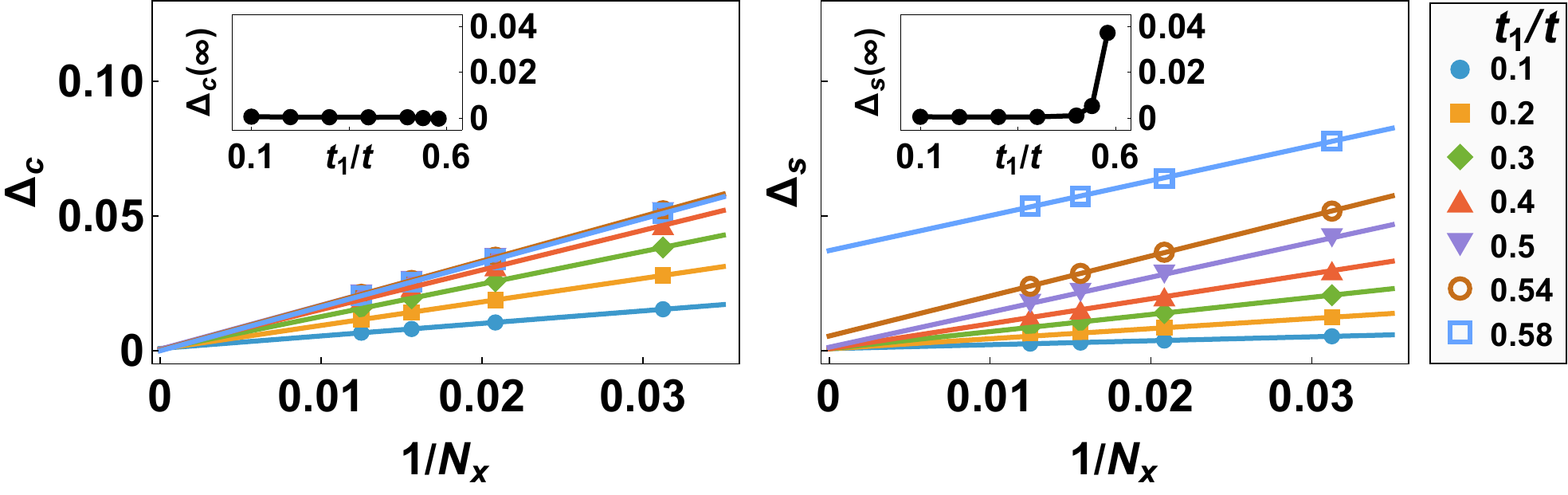}
  \caption{Scalings of the charge gap $\Delta_c$ and spin gap $\Delta_s$ with $1/N_x$. The insets show the variations of extrapolated gaps with $t_1/t$ in the thermodynamic limit. After the transition, the spin gap becomes finite while the charge gap remains zero. The truncation error is fixed at $\epsilon = 1.0\times 10^{-7}$.}\label{fig:gap}
\end{figure}

\subsection{Energy gap}
The phase transition can also be identified from charge and spin gaps, defined respectively as \cite{pengGapless2021}
\begin{align}
&\Delta_c = \frac{1}{2}\left(E_{\frac{N_e}{2}+1,\frac{N_e}{2}+1}+E_{\frac{N_e}{2}-1,\frac{N_e}{2}-1}-2E_{\frac{N_e}{2},\frac{N_e}{2}}\right), \nonumber \\
&\Delta_s = E_{\frac{N_e}{2}+1,\frac{N_e}{2}-1}-E_{\frac{N_e}{2},\frac{N_e}{2}},
\end{align}
where $E_{N_\uparrow,N_\downarrow}$ denotes the ground-state energy in the sector with total electron number $N_e = N_\uparrow+N_\downarrow$ and total $z$ component of spin $S_z = (N_\uparrow-N_\downarrow)/2$. Figure \ref{fig:gap} shows linear scaling of the gap size with $1/N_x$, with the insets displaying the extrapolated values in the thermodynamic limit. For $t_1<0.5t$, both the charge and spin gaps vanish, consistent with algebraically decaying charge and spin correlation functions in this weak intralayer hopping regime. When $t_1>0.5t$, a finite spin gap opens, which explains the absence of quasi-long-range SDW order.

\subsection{Central charge}

\begin{figure}
  \includegraphics[width = 0.48\textwidth]{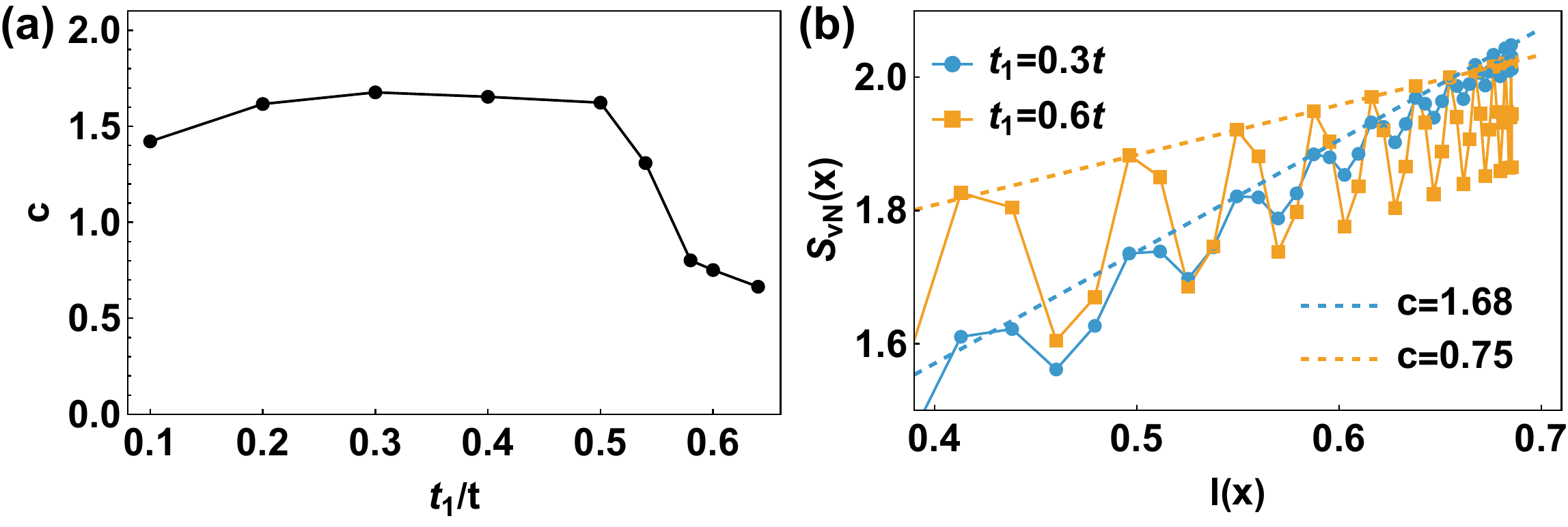}
  \caption {(a) The evolution of central charge $c$ with $t_1/t$. (b) The scaling of von Neumann entanglement entropy $S_{\text{vN}}$ with subsystem length $x$, with the horizontal axis labeling $l(x) = \ln \left(\frac{2N_x}{\pi}\sin \frac{\pi x}{N_x}\right)/6$. The central charge is given by the slope of the fitting curves (dashed lines). }\label{fig:centralCharge}
\end{figure}

The evolution of energy gap suggests that the system supports gapless charge and spin modes in the SDW phase, and only the charge mode in the LE phase. To determine the number of such gapless modes, we may extract the central charge, denoted by $c$, following the scaling relation of von Neumann entanglement entropy $S_{\text{vN}}$ with subsystem length $x$, which is given by \cite{calabrese2004}
\begin{equation}
  S_{\text{vN}}(x) = \frac{c}{6}\ln \left(\frac{2N_x}{\pi}\sin \frac{\pi x}{N_x}\right) + S_0,
\end{equation}
with $S_0$ being constant for a given system. In the quasi-1D system considered here, central charge simply counts the total number of gapless spin and charge modes. As shown in Fig. \ref{fig:centralCharge}(a), $c$ drops by approximately 1 across $t_1/t=0.5$, verifying our previous claim that one gapless spin mode is gapped out at the transition point. With the current numerical precision (truncation error $\epsilon\geq 1.0\times 10^{-7}$), we cannot accurately determine the central charge, which are expected to be integers in this system. The quality of fitting is also affected by the additional oscillatory decaying terms in the entanglement entropy that arise due to open boundaries \cite{laflorencie2006,calabrese2010}. In Supplemental Material \cite{supplemental}, we show the extracted central charge for $t_1/t=0.1$ under periodic boundary conditions, which are greater than 2. In combination with the results obtained under OBC, our best estimate is that the SDW phase supports one gapless mode in both charge and spin sectors ($c=2$), while the LE phase hosts a single gapless mode only in the charge sector ($c=1$), which is a well-known feature of LE liquid.

\subsection{Pairing symmetry}

\begin{figure}[t]
  \includegraphics[width = 0.48\textwidth]{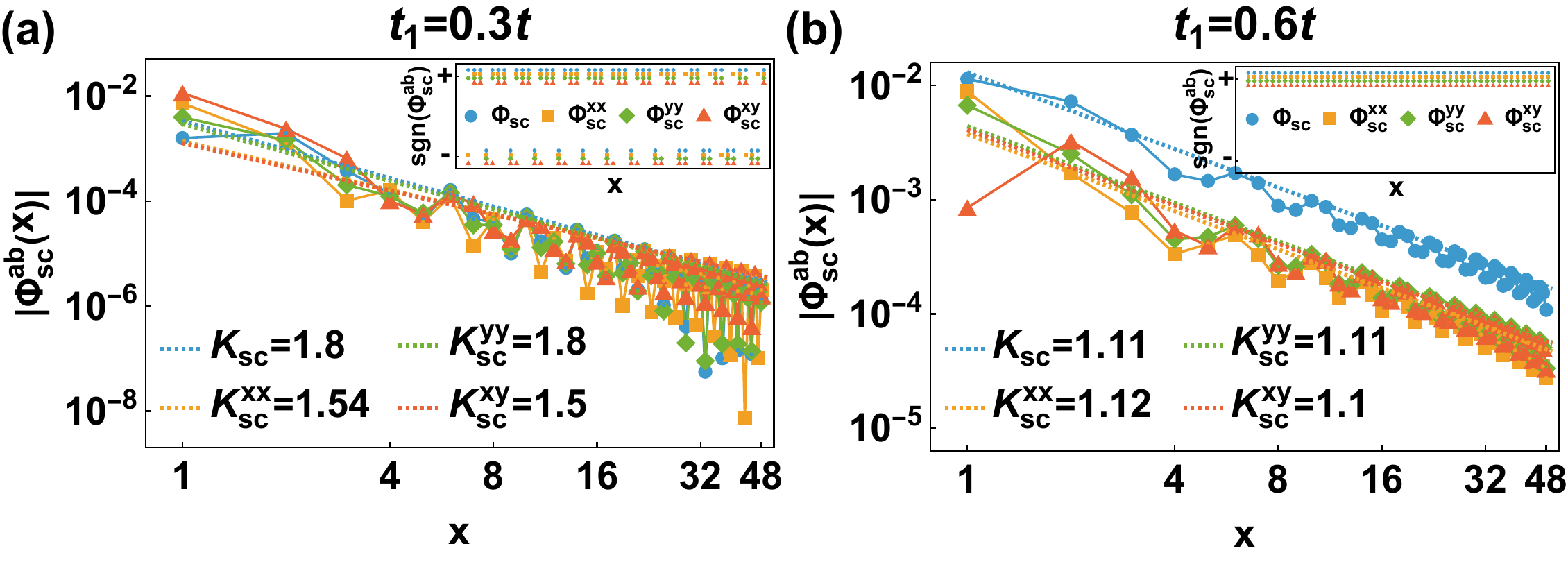}
  \caption {Pairing correlations of intralayer and interlayer singlets in the two phases. All the correlation functions decay algebraically, with $K_{sc}$ denoting corresponding decay exponents. (a) In the SDW phase, the sign of correlation functions show strong oscillations with period 4, as shown in the inset. (b) In the LE phase, the four correlation functions remain positive and exhibit similar decay exponents. The correlation of interlayer singlets is relatively stronger than that of intralayer ones.}\label{fig:pairing}
\end{figure}

Previously, we only consider the pairing correlation between interlayer singlets, which is dramatically enhanced and shows weak oscillations when the system is driven into the LE phase. If such pairing develops into long-range order in the 2D limit, the superconducting phase will exhibit $s_\pm$ symmetry, with the pairing order parameter on the bonding and anti-bonding bands showing $\pi$-phase difference. This can be seen from the Fourier transformation of interlayer pairing operator
\begin{equation}
  \sum_{\vec r} (c_{\vec r,1\uparrow} c_{\vec r,2\downarrow}-c_{\vec r,1\downarrow}c_{\vec r,2\uparrow}) = \sum_{\vec k} \cos k_z c_{\vec k,\uparrow} c_{-\vec k,\downarrow},
\end{equation}
with $c_{\vec k,\sigma} = \sum_{\vec r,\lambda} c_{\vec r,\lambda\sigma} e^{-i(\vec k\cdot \vec r + k_z\lambda)}/\sqrt{N}$. In addition to the interlayer singlet pairing, we also consider the correlations between intralayer singlets, defined by
\begin{equation}
  \Phi_{sc}^{ab}(x) = \frac{1}{2}\sum_\lambda \langle\Delta_{a,\lambda}^\dagger(\vec r_0) \Delta_{b,\lambda}(\vec r_0+x\vec e_x)\rangle,
\end{equation}
where $\Delta_{a,\lambda}(\vec r) = (c_{\vec r,\lambda\uparrow}c_{\vec r+\vec e_a,\lambda\downarrow} - c_{\vec r,\lambda\downarrow}c_{\vec r+\vec e_a,\lambda\uparrow})/\sqrt{2}$, denoting singlet pairing between nearest neighbors along $a \ (=x,y)$ direction. We plot the pairing correlations $\Phi_{sc}^{xx}$, $\Phi_{sc}^{yy}$ and $\Phi_{sc}^{xy}$ in Fig. \ref{fig:pairing}, together with $\Phi_{sc}$ defined in Eq.(\ref{eq:correlation}). In the SDW phase, $\Phi_{sc}^{xx}$ and $\Phi_{sc}^{xy}$ decay slightly slower than $\Phi_{sc}^{yy}$ and $\Phi_{sc}$, and their signs change periodically, as shown in Fig. \ref{fig:pairing}(a). This is in contrast to the LE phase where none of the pairing correlations demonstrate sign oscillation. In this phase, the interlayer singlet pairing correlation is stronger than intralayer singlet correlations, although their decay exponents are very close. We note that, as shown in the inset of Fig. \ref{fig:pairing} (b), $\Phi_{sc}^{xy}$ is always positive, suggesting that intralayer pairing along $x$ and $y$ directions should take the same sign. Considering the $C_4$ symmetry in the 2D limit, we expect $\langle\Delta_{x,\lambda}\rangle=\langle\Delta_{y,\lambda}\rangle$ if they could develop into truly long-ranged orders. From the Fourier transformation of the pairing operator $\Delta_{a,\lambda}(\vec r)$ in the momentum space, given by 
\begin{equation}
  \sum_{\vec r,\lambda}\Delta_{a,\lambda}(\vec r) = \sum_{\vec k}  \sqrt{2}\cos k_a c_{\vec k,\uparrow} c_{-\vec k,\downarrow},
\end{equation}
we can readily conclude that the intralayer pairing exhibits extended $s$-wave symmetry. Therefore, in the parameter regime we consider, the LE phase favors extended $s$-wave superconductivity, with the pairing order parameter in the bonding and anti-bonding band exhibiting $\pi$-phase difference. 

\section{Discussion and conclusion}

The quantum phase transition driven by increasing ratio of intralayer to interlayer hopping separates SDW and LE phases, signifying the suppression of SDW and the enhancement of superconductivity. We may understand the phase diagram starting from the dimer limit, where the degenerate ground states are superpositions of the singlet and bonding dimers. Increasing intralayer hopping $t_1$ enhances phase coherence among interlayer singlets and thus strengthens superconductivity. However, this enhancement is hindered by increasing involvement of other types of dimers shown in Table \ref{table:dimer}, as the $t_1$ term favors the coupling among different types of dimers. Hence, a slight upturn in the decay exponent curve of pairing correlations is observed at sufficiently large $t_1$. The oscillation of correlation functions in the weak intralayer hopping regime could be attributed to the different properties of singlet and bonding dimers in spin, charge and pairing, and we find that SDW becomes the leading (quasi-long-range) order. Its disappearance in the larger $t_1/t$ regime could possibly arise from the increasing involvement of other types of dimers, in which case this specific spin-density modulation is no longer favored.

Given that the intralayer hopping between the $d_{3z^2-r^2}$ orbitals in bilayer nickelates is weak, with $t_1\approx 0.2 t$ \cite{luoBilayer2023}, our results suggest that this orbital alone is insufficient to support high-temperature superconductivity. However, moderate hybridization with the more itinerant $d_{x^2-y^2}$ orbitals could strengthen superconductivity by increasing phase coherence among preformed Cooper pairs \cite{kivelsonMaking2002,bergRoute2008,yangInterlayer2023}. Further verification of this scenario requires treating both orbitals on equal footing \cite{shenNumerical2025} and comparing the resulting phase transitions with those found in the effective single-orbital model. The substantial enhancement of superconductivity near the transition point may help explain the experimental observations that the superconducting transition temperature increases sharply upon reaching the critical pressure \cite{sunSignatures2023,liIdentification2025}, as both intralayer and interlayer hopping amplitudes are pressure-dependent. 

The absence of CDW in the weak intralayer hopping regime suggests that the proposed CDW in bilayer nickelates at ambient pressure may not originate from the $d_{3z^2-r^2}$ orbital. A recent experiment on La$_3$Ni$_2$O$_7$ \cite{khasanovOxygenisotope2025} observed a pronounced isotope effect in the CDW transition temperature, indicating that electron-phonon coupling likely contributes to the formation of CDW order. In contrast, the transition temperature of SDW appears insensitive to isotope substitution, supporting its purely electronic origin. In the relatively strong intralayer hopping regime, we reveal an intense competition between CDW and superconductivity at the doping level considered. In bilayer nickelates, the filling of the $d_{3z^2-r^2}$ orbital can be influenced by hybridization with the $d_{x^2-y^2}$ orbital, the onsite energy difference between the two orbitals, extra doping from substrates \cite{zhouAmbientpressure2025,yueCorrelated2025}, and even van der Waals interactions between neighboring bilayers \cite{pengImportance2024}. Given the possible doping dependence of superconducting transition temperature in the bilayer Hubbard model \cite{liuHybridizationinduced2025,nomuraStrongcoupling2025}, further study is warranted to systematically explore how doping influences the interplay between CDW and superconductivity in the context of bilayer nickelates.

In summary, we identify a quantum phase transition driven by increasing the ratio of intralayer to interlayer hopping, within a single-orbital framework that explicitly includes only the Ni-$d_{3z^2-r^2}$ orbitals of bilayer nickelates. The transition is characterized by the disappearance of SDW, the emergence of CDW, and a pronounced enhancement of superconductivity.  We show that $d_{3z^2-r^2}$ electrons could develop superconducting order provided the effective intralayer hopping is sufficiently strong. In the scenario where the $d_{x^2-y^2}$ electrons play the active role in pairing, this result implies that the superconducting transition temperature may be further elevated by the condensation of $d_{3z^2-r^2}$ electrons. Our study demonstrates that the investigation of quantum phase transitions can offer valuable insights into the competition between superconducting and density-wave orders in bilayer nickelates, and may help guide future efforts to optimize superconductivity in this platform.

\begin{acknowledgments}
  
  This work was supported by National Natural Science Foundation of China (Grant Nos. 12488101, 12374458, 11974323, and 12474134), Innovation Program for Quantum Science and Technology (Grant No. 2021ZD0302800), Strategic Priority Research Program of Chinese Academy of Sciences (Grant No. XDB0510200), and Anhui Provincial Natural Science Foundation (Grant No. 2508085MA004).

\end{acknowledgments}

The data that support the findings of this article are openly available \cite{dataset}.

\bibliography{bilayer}

\onecolumngrid
\clearpage

\setcounter{equation}{0}
\setcounter{figure}{0}
\setcounter{page}{1}
\renewcommand{\theequation}{S\arabic{equation}}
\renewcommand{\thefigure}{S\arabic{figure}}

\begin{center}
    {\textbf{\large Supplemental Material for ``Quantum phase transition driven by competing intralayer \\[0.2cm] and interlayer hopping in bilayer nickelates"}\\[1cm]}
\end{center}

\appendix

In the Supplemental Material, we present how the local densities and correlation functions vary with truncation errors, detail the procedure for extracting decay exponents from charge-density profiles, discuss the extension to a wider bilayer system with $N_y=4$, and show the central charge extracted under periodic boundary conditions.

\section{Truncation error}

The truncation error $\epsilon$ is a key parameter in DMRG simulations. It is defined as the sum of eigenvalues of disregarded states in the reduced density matrix when the system block grows during the sweeps. A smaller value of $\epsilon$ indicates fewer states omitted, and thus higher numerical accuracy. In our simulations, we fix $\epsilon$ at different values for each run and investigate how the resulting local densities and correlations vary with $\epsilon$. We then extrapolate these quantities to the limit $\epsilon = 0$ using quadratic fit, providing an estimate of the corresponding properties of the true ground state. The smallest truncation error considered is $\epsilon = 1.0\times 10^{-7}$.

\begin{figure*}[h]
  \includegraphics[width = 0.98\textwidth]{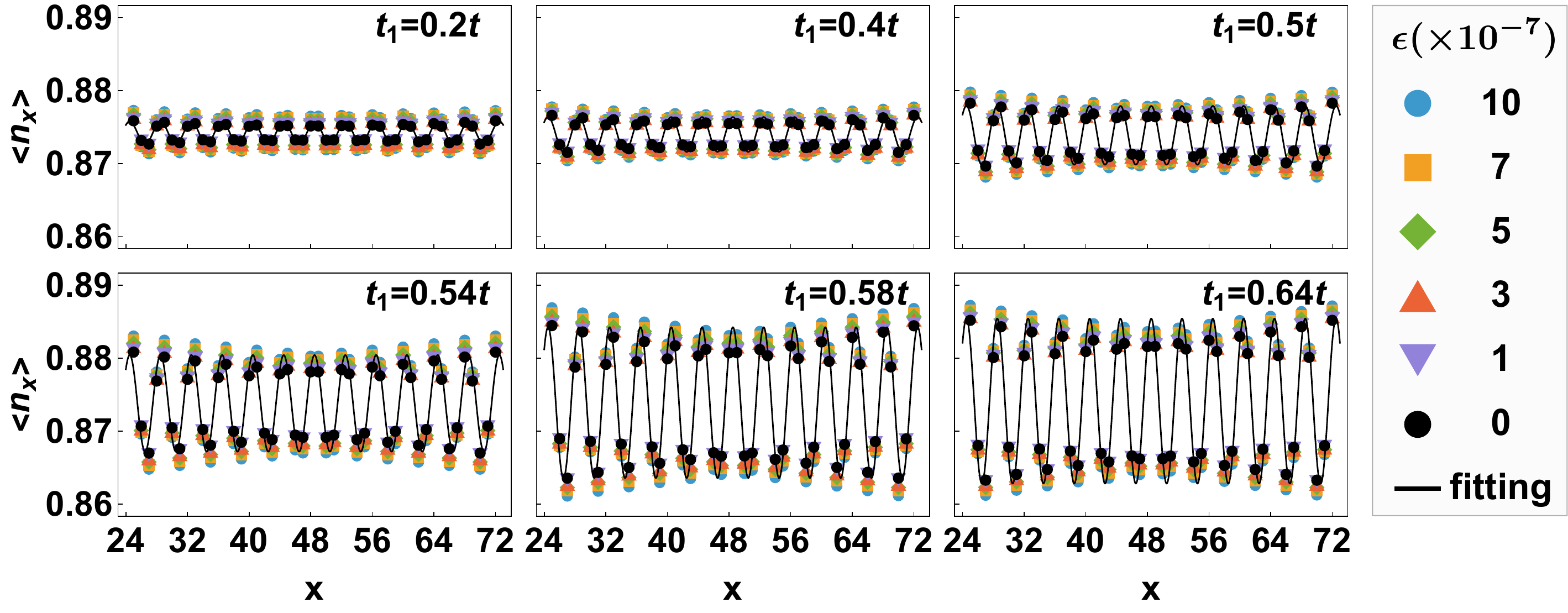}
  \caption{Evolution of charge-density profile with truncation error for various $t_1$. The circles in black represent the extrapolated values at $\epsilon=0$. The lines indicate fitting curves of Eq.(\ref{eqS:density_fitting}). In all the figures, $N_x=96$.}\label{figS:density_error}
\end{figure*}

\begin{figure*}[h]
  \includegraphics[width = 0.98\textwidth]{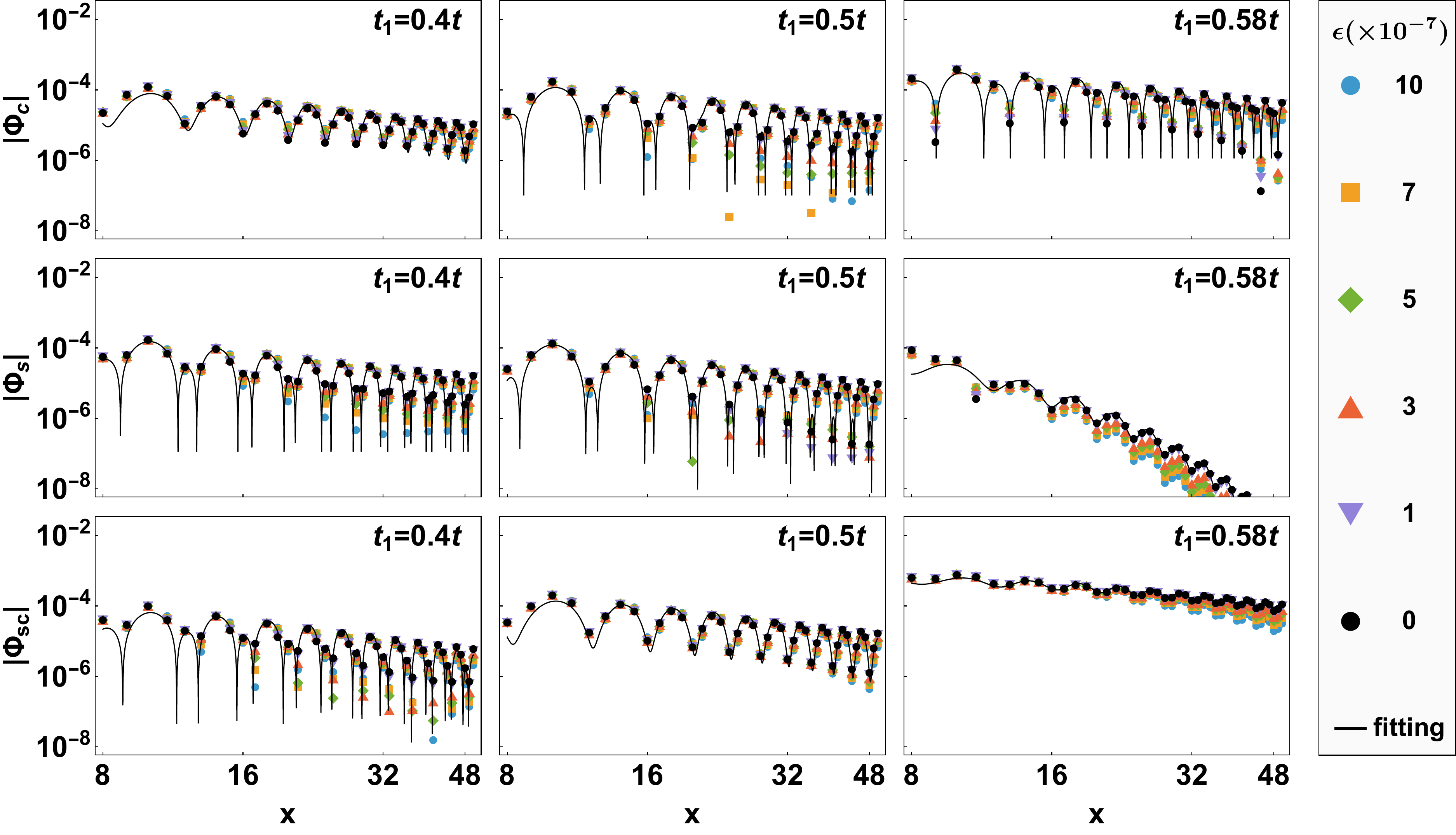}
  \caption{Evolution of correlation functions with truncation error for three representative values of $t_1/t$ plotted in the log-log scale. The black circles represent extrapolated values and the lines indicate corresponding fitting curves, either $|\Phi(x)|= C x^{-K} [1+B\cos(\pi x/2+\theta)]$, or $|\Phi(x)|= C \exp^{-x/\xi} [1+B\cos(\pi x/2+\theta)]$ (for $|\Phi_s|$ at $t_1=0.58t$). }\label{figS:corr_error}
\end{figure*}

As shown in Fig. \ref{figS:density_error}, a relatively large truncation error tends to overestimate the oscillation amplitude in the charge-density profile. In contrast, DMRG typically underestimates the long-distance correlations, particularly in critical systems, where correlation length diverges. To observe the expected power-law decay of correlation functions in a gapless system with finite size, it is necessary to use sufficiently small truncation errors. As shown in Fig. \ref{figS:corr_error}, the power-law behavior becomes more obvious as the truncation error is reduced.

\section{Decay exponents from charge-density profile}

In systems with open boundaries, the ground-state charge density may exhibit oscillations. However, this does not necessarily imply the presence of charge density wave (CDW) order, as such oscillations can also arise purely from Friedel oscillations \cite{whiteFriedel2002}. To distinguish between these two scenarios, one can analyze the decay exponent $K_c$ extracted from the charge-density profile. If the exponent is close to or greater than 2, it is likely that true CDW order is absent in the 2D limit.

\begin{figure*}[ht]
  \includegraphics[width = 0.98\textwidth]{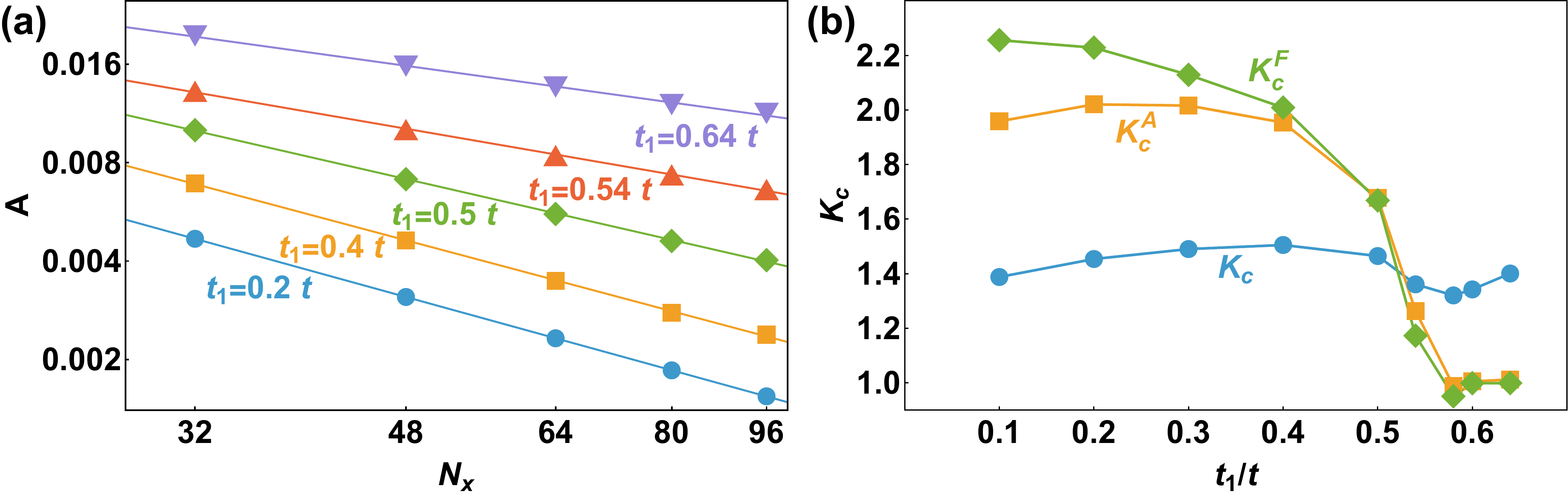}
  \caption{(a) Scaling of oscillation amplitude with system length. Lines indicate the linear-fitting curves plotted in the log-log scale, and the slopes represent the decay exponent $K_c^A$. (b) The comparison of decay exponents obtained using different methods. $K_c^A$ and $K_c^F$ are extracted from charge-density oscillations, and $K_c$ is obtained from the charge-charge correlation. }\label{figS:exponentDensity}
\end{figure*}

We compare the results obtained from two different methods. In the first method, we fit the density profile near the center of the system using the ansatz \cite{jiangSuperconductivity2019}
\begin{align}
  \langle n_x \rangle = n_0+A\cos(k_0 x-\phi),\label{eqS:density_fitting}
\end{align}
with $k_0=\pi/2$ for $1/8$-doping and $\phi=\pi(N_x+1)/4$, as shown in Fig. \ref{figS:density_error}. The fitted amplitude $A$ is expected to follow a power-law relation with the system size $N_x$, \emph{i.e.}, $A \sim  (N_x)^{-K_c^A/2}$, as demonstrated in Fig. \ref{figS:exponentDensity} (a), where $K_c^A$ is the decay exponent. 

Alternatively, the density oscillation can be directly fitted using the Friedel oscillation formula \cite{whiteFriedel2002}:
\begin{align}
  \langle n_x \rangle = n_0 + \frac{\tilde A \cos (k_0 x - \phi)}{[N_x \sin (\frac{\pi x}{N_x})]^{K_c^F/2}},
\end{align}
with $k_0$ and $\phi$ taken to be the same as in Eq.(\ref{eqS:density_fitting}). Figure \ref{figS:exponentDensity}(b) compares the decay exponents $K_c^A$ and $K_c^F$ obtained from these two approaches, along with $K_c$ extracted from the charge-charge correlation functions. In the regime of weak intralayer hopping, both $K_c^A$ and $K_c^F$ indicate weak charge-density oscillations, although their precise values differ slightly. The fact that $K_c$, the decay exponent of charge correlation, is smaller than 2 suggests the existence of gapless charge mode. Across the transition at $t_1=0.5t$, both $K_c^A$ and $K_c^F$ drop sharply and become comparable, suggesting that the density oscillation is enhanced in this regime and that the system develops quasi-long-range CDW order.

\section{$N_y = 4$}

In the main text, we present results only for $N_y = 2$. At fixed truncation error, the number of states required grows exponentially with the system width $N_y$. Here, we show the density oscillations and correlation functions for $t_1 = 0.2t$ in a $32 \times 4$ bilayer system. Due to the rapidly increasing computational cost with increasing $t_1/t$, we are unable to access the regime beyond the transition point. For $N_y = 4$, we impose periodic boundary conditions in the $y$-direction.

\begin{figure*}[ht]
  \includegraphics[width = 0.98\textwidth]{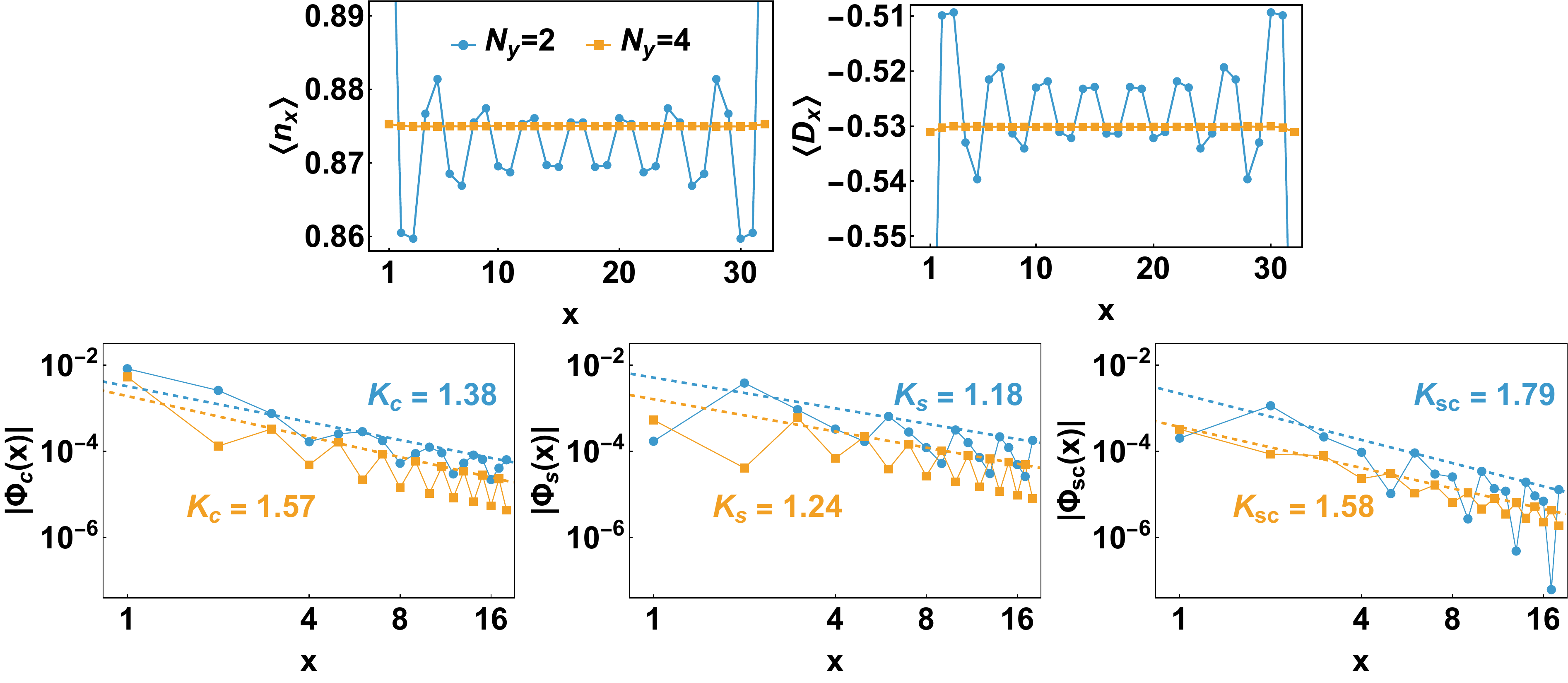}
  \caption{Comparison between systems with $N_y=2$ and $N_y=4$ for $t_1=0.2t$. In both cases, $N_x=32$ and the truncation error is fixed at $\epsilon = 1.0\times 10^{-7}$. The lower panels show correlations in log-log scale, wherein the dashed lines are power-law fitting curves, and $K$ represent the decay exponent. }\label{figS:ny4}
\end{figure*}

As shown in Fig. \ref{figS:ny4}, the charge density and spin correlations along the rungs barely oscillate in the $N_y=4$ case. This observation supports our earlier argument that density oscillations remain weak before $t_1/t$ reaches the transition point, and no CDW is present. In this regime, the charge, spin, and pairing correlations all display power-law decay, with the oscillation period reduced to half of that in the $N_y = 2$ case. Notably, the spin-spin correlation remains the strongest among them, indicating that quasi-long-range SDW order continues to dominate, similar to the $N_y=2$ case.

\section{central charge in periodic boundary condition}

\begin{figure*}
  \includegraphics[width = 0.98\textwidth]{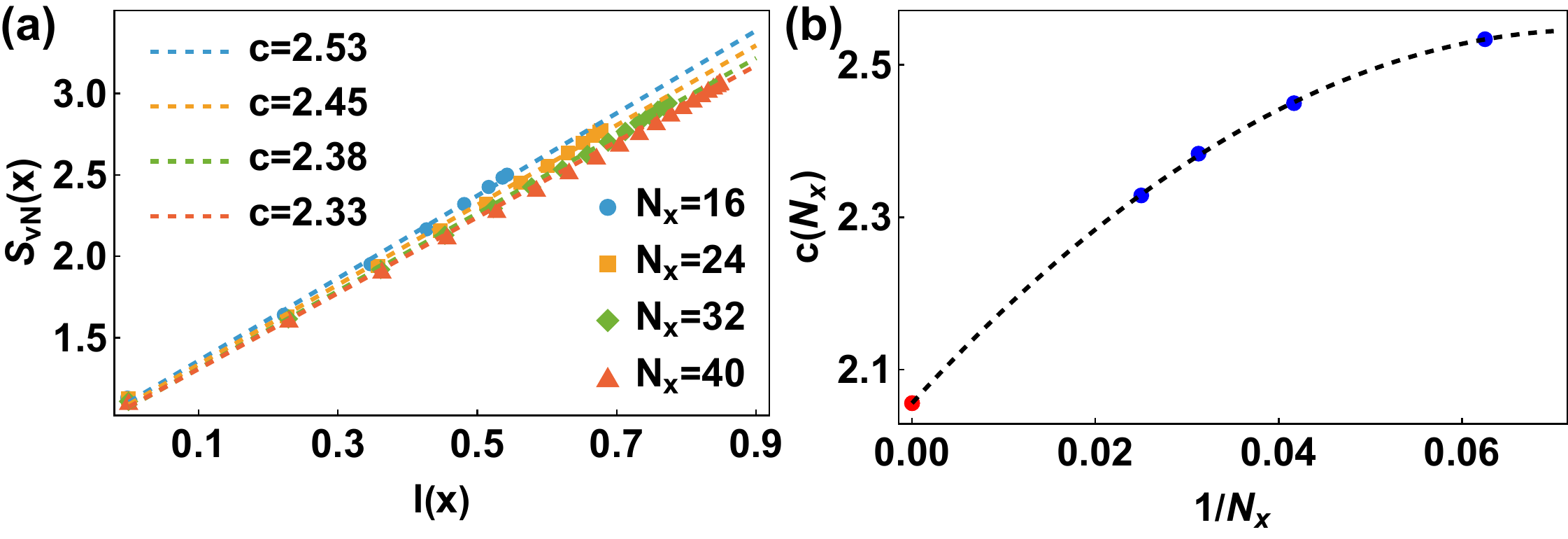}
  \caption{(a) Entanglement entropy at $t_1/t=0.1$ for various system size up to $N_x=40$. The horizontal axis label $l(x) = \log \left(\frac{N_x}{\pi}\sin \frac{\pi x}{N_x}\right)/3$. (b) Quadratic fitting of the central charge as a function of system size. The red point indicates the extrapolated central charge in the thermodynamic limit. The truncation error is fixed at $\epsilon = 1.0\times 10^{-7}$.}\label{figS:centralChargePBC}
\end{figure*}

As mentioned in the main text, with open boundary conditions (OBC), the entanglement entropy admits additional oscillating terms that decay from the boundaries. Here, we present entropy plot under periodic boundary conditions (PBC) for $t_1/t=0.1$, as shown in Fig. \ref{figS:centralChargePBC}(a). The scaling relation of entanglement entropy with subsystem size takes the following form
\begin{equation}
  S_{\text{vN}}(x) = \frac{c}{3}\log \left(\frac{N_x}{\pi}\sin \frac{\pi x}{N_x}\right) + S_0.
\end{equation}
The oscillating terms do not appear in this case due to the absence of boundaries. In contrast to the results in open-boundary systems, the central charge extracted under PBC is greater than 2, and decreases when the system size $N_x$ is increased. Hence, we performed a quadratic fitting based on the limited data points, as shown in Fig. \ref{figS:centralChargePBC}(b), and the extrapolated central charge in the thermodynamic limit ($N_x\rightarrow \infty$) is very close to $2$. Combined with central charge obtained under OBC, we estimate $c=2$ for the SDW phase, corresponding to two gapless modes.

\end{document}